\documentclass[prb,letterpaper,aps,floatfix,twocolumn]{revtex4}
\usepackage{graphicx}
\usepackage{amsmath}
\newcommand{\beq}{\begin{equation}}
\newcommand{\eeq}{\end{equation}}
\newcommand{\bk}{{{\bf{k}}}}

\newcommand{\br}{{{\bf{r}}}}
\newcommand{\bR}{{{\bf{R}}}}

\newcommand{\bG}{{{\bf{G}}}}
\newcommand{\bA}{{\bf{A}}}
\newcommand{\bB}{{\bf{B}}}
\newcommand{\ba}{{\bf{a}}}

\newcommand{\bq}{{\bf{q}}}

\newcommand{\bp}{{\bf{p}}}
\newcommand{\bb}{{\bf{b}}}
\newcommand{\bm}{{\bf{m}}}

\newcommand{\beqa}{\begin{eqnarray}}
\newcommand{\eeqa}{\end{eqnarray}}
\newcommand{\ra}{\rangle}
\newcommand{\la}{\langle}

\newcommand{\dg}{{\dag}}
\newcommand{\pdg}{{\vphantom\dag}}
\begin{document}
\title{Fractional quantum Hall effect and featureless Mott insulators}
\author{A.A. Burkov}
\affiliation{Department of Physics and Astronomy, University of Waterloo, Waterloo, Ontario, 
Canada N2L 3G1}
\date{\today}
\begin{abstract}
We point out and explicitly demonstrate a close connection that exists between featureless 
Mott insulators and fractional quantum Hall liquids.
Using {\em magnetic Wannier states} as the single-particle basis in the lowest 
Landau level (LLL), we demonstrate that the Hamiltonian of interacting bosons 
in the LLL maps onto a Hamiltonian of a featureless 
Mott insulator on triangular lattice, formed by the magnetic Wannier states.
The Hamiltonian is remarkably simple and consists only of short-range repulsion and 
ring-exchange terms.     
\end{abstract}
\maketitle
\section{Introduction}
The goal of this work is to explicitly demonstrate a close connection that exists
between two paradigmatic strongly correlated systems: a Mott insulator and a fractional 
quantum Hall liquid (FQHL). 
The connection is to some degree almost obvious.
Fractional quantum Hall effect (FQHE) arises when a two-dimensional (2D) liquid of interacting charged quantum particles (either fermions or bosons), placed in a perpendicular magnetic field, 
becomes incompressible at certain commensurate filling factors, i.e. ratios of the number of 
particles to the number of available degenerate single-particle states in lowest Landau level (LLL), 
which is equal to the number of magnetic flux quanta piercing the sample. 
Mott insulator is a very similar thing: an incompressible state arising at specific filling factors, 
in this case given by the ratio of the number of particles to the number of available 
degenerate localized Wannier orbitals in a given crystal lattice. 
An important difference between a FQHL and a Mott insulator is that, while the FQHL 
is a {\em liquid}, i.e. is featureless and does not break any symmetries, a Mott insulator can be 
either a {\em liquid} or a {\em crystal}, i.e. either be featureless or break the underlying lattice symmetry.  
In fact, in most cases, at a general fractional filling factor (for bosons the filling factor is defined 
here as the ratio of the number of particles to the number of orbitals, for electrons it is half that ratio), 
a Mott insulator will break symmetry, as happens, for example, in the parent compounds of the cuprate 
superconductors.~\cite{Lee-Wen} 
The connection is thus between a FQHL and a {\em featureless Mott insulator}. 
This, in our opinion, is the main point that makes this connection interesting. 
Featureless Mott insulators have been actively searched for in recent years, both experimentally 
and theoretically.\cite{Wen} Even though a lot of progress has been made, in particular concrete microscopic models
with featureless Mott insulator ground states have been proposed, \cite{Motrunich} the general ingredients, which are necessary
in a microscopic model to obtain a featureless Mott insulator ground state, are not yet known. 
We believe that the FQHL connection may prove to be a useful contribution to this field.   

While (at least superficially) rather obvious, the FQHL to Mott insulator connection has been largely unexplored.  
Only very recently it was explicitly pointed out in a series of papers by D.-H. Lee {\it et al.} \cite{Lee} and by Bergholtz and Karlhede. \cite{Karlhede08}
It was demonstrated in these works that for a quantum Hall system on a torus there exists a limit, namely the quasi-one-dimensional (quasi 1D)
limit, reached when one of the dimensions of the torus is made comparable to or even smaller than the magnetic length, in which  
the Mott insulator connection becomes simple and explicit and the fractional quantum Hall liquid becomes a simple crystal (not a featureless Mott insulator), with Landau-orbital 
positions playing the role of the ``lattice sites". 
It was further demonstrated that the evolution from the quasi 1D to the physical 2D limit is (in many cases) smooth, with the 2D 
fractional quantum Hall liquid ground state inheriting the discrete degeneracy of the 1D crystal, 
but in the form of a topological degeneracy, as the fractional quantum Hall liquid is featureless. 

While very elegant and appealing, the picture of Refs.~[\onlinecite{Lee,Karlhede08}] has some imperfections. The first one is that, while the evolution 
from the quasi 1D to the 2D limit may be smooth, the 2D thermodynamic limit is still singular in the sense that the 2D fractional quantum Hall liquid is 
certainly not a simple crystal with a broken translational symmetry that obtains in the quasi 1D limit, but is a featureless liquid with topological order.
Moreover, the evolution from the 1D to the 2D limits is in fact not always smooth: for example 
it is not smooth in the case of the $\nu=1/2$ composite fermion Fermi liquid state.~\cite{Karlhede08}

It would be more satisfying to have an approach that could establish the Mott insulator to FQHL connection directly in two dimensions. The main obstacle here is a problem with notation. Namely, the standard choices for LLL orbital eigenstates, like e.g. Landau-gauge orbitals used in Refs.~[\onlinecite{Lee,Karlhede08}], are delocalized. 
This means that the energy cost for doubly occupying such orbitals vanishes in the 2D thermodynamic limit. Then it becomes hard make an analogy to Mott insulator, since the Mott insulator physics is most 
easily described in terms of prohibiting double occupation of some orbitals or 
nearest-neighbor groups of orbitals. This physics is completely obscured if one uses 
delocalized states as single particle basis. 
What is needed to make the FQHL to Mott insulator connection is a single-particle 
LLL basis, that would consists of functions, localized in all directions in the 2D plane, analogous 
to Wannier functions in insulators. 
This appears to be problematic. 
It is well-known \cite{Thouless84} that constructing exponentially localized Wannier 
orbitals in the LLL is impossible: exponential localization and a nonvanishing 
topological invariant, the Chern number, which characterizes Landau levels and which is the source of 
the precisely quantized Hall conductance, are incompatible. 
For our purposes, however, exponential localization is unnecessary: all we need is a basis 
of normalizable orbitals, which have a finite energy cost of double occupation. 
It was explicitly demonstrated by Rashba {\it et al.} \cite{Rashba}, that it is in fact possible to construct
exactly such a basis of quasilocalized Wannier-like orbitals, called {\em magnetic Wannier functions} in 
Ref.~[\onlinecite{Rashba}]: these wavefunctions have a gaussian core and a $1/r^2$ tail, 
$1/r^2$ being the fastest decay compatible with a nontrivial Chern number.  
While not exponentially localized, these 
magnetic Wannier orbitals are normalizable and have a finite energy cost of double occupation. 
It will be demonstrated in this paper that using magnetic Wannier functions as a single-particle basis in the LLL, 
it is possible to map the problem of interacting particles in 2D in a strong magnetic field onto a problem of 
interacting particles on triangular lattice with one magnetic flux quantum per unit cell, described by a short-range time-reversal 
invariant Hamiltonian. 
While we believe that the final result is valid, with possible minor modifications, for either bosons or fermions, 
the arguments, leading to this result, work only for bosons. We will thus focus henceforth on the FQHE of charged bosons. 
The above mapping then implies that at the filling factors, at which the bosons in the LLL exhibit FQHE, the ground state 
of the equivalent model on the triangular lattice is a featureless Mott insulator with topological order. 

The rest of the paper is organized as follows. In section \ref{sec:2} we review, for reader's convenience, the construction 
of the magnetic Wannier basis and point out why its straightforward application to our problem is nontrivial.  
In section \ref{sec:3} it is demonstrated that the magnetic Wannier basis Hamiltonian has an emergent low-energy 
long-wavelength symmetry, that drastically reduces the number of terms in the Hamiltonian and makes it possible 
to construct a simple short-range lattice Hamiltonian, faithfully representing bosons in the LLL. 
In section \ref{sec:4} we explicitly discuss the physical properties of this lattice Hamiltonian and we conclude 
with a brief summary of the results in section \ref{sec:5}. 

\section{Magnetic Wannier basis}
\label{sec:2}
We will start by reviewing, for reader's convenience, the construction of the magnetic Wannier basis,
proposed in Ref.[\onlinecite{Rashba}]. 
One starts from the zero-angular-momentum symmetric gauge LLL wavefunction:
\beq
\label{eq:1} 
c_0(\br) = \frac{1}{\sqrt{2 \pi \ell^2}} e^{-r^2/4 \ell^2},
\eeq 
where $\ell$ is the magnetic length. This wavefunction has the form of an atomic-like orbital, 
centered at the origin. 
To construct a complete set of such atomic-like orbitals in the LLL, one can translate the above 
wavefunction, using magnetic translation operators, to sites of any 2D Bravais lattice.
A priori, the only restriction one can place on the form of this lattice is that 
the unit cell must contain exactly one magnetic flux quantum, or in other words, its area must be equal 
to $2 \pi \ell^2$. 
However, it will be demonstrated below that in fact the most natural choice is the triangular lattice. 
Translating $c_0(\br)$ to sites of this triangular lattice, we obtain:
\beqa
\label{eq:2}
c_\bm(\br)&=&T_{m_1 \ba_1} T_{m_2 \ba_2} c_0(\br)  \nonumber \\
&=&\frac{(-1)^{m_1 m_2}}{\sqrt{2 \pi \ell^2}} e^{-(\br - \br_\bm)^2/ 4 \ell^2 + (i/ 2 \ell^2) \hat z \cdot (\br \times \br_\bm)},
\eeqa
where $\ba_1 = a \hat x, \,\, \ba_2 = a (\hat x + \sqrt{3} \hat y)/2$ are the basis 
vectors of the triangular lattice, $\bm = (m_1, m_2)$ with integer $m_i$ label 
the lattice sites and $T_\bR = \exp[-i \bR \cdot (\bp - e \bA/c)]$ are the magnetic translation operators 
(we will take the charge of the bosons to be $-e$ and assume the symmetric gauge $\bA = \frac{1}{2} \bB \times \br$). 
The requirement that the unit cell contain exactly one magnetic flux quantum gives $|\ba_1 \times \ba_2| = 2 \pi \ell^2$, 
fixing the lattice constant in our case to be $a = \sqrt{4 \pi \ell^2 / \sqrt{3}}$. 

The set of functions $c_{\bm}(\br)$ looks similar to a complete but nonorthogonal set of atomic orbitals in a crystal. 
However, this appearance is deceptive, since this set of functions in fact possesses a very nontrivial property that makes it 
very different from a simple set of localized atomic orbitals. This property is embodied in the following
identity, first established by Perelomov: \cite{Perelomov}
\beq
\label{eq:3}
\sum _{\bm} (-1)^{m_1+m_2} c_\bm(\br) = 0. 
\eeq
Eq.(\ref{eq:3}) means that the set of functions $c_{\bm}(\br)$ is in fact {\em overcomplete} by exactly one 
state. This identity is the origin of the nontrivial topological properties of the {\em magnetic Bloch states}, which we will 
construct below as linear combinations of $c_{\bm}(\br)$, namely the nontrivial Chern number characterizing the LLL. 
It will also play a very important role in our analysis. 

Given the set of atomic-like wavefunctions $c_{\bm}(\br)$, one can follow the standard 
procedure to construct magnetic Wannier functions. One first constructs Bloch 
functions out of linear combinations of $c_{\bm}(\br)$ as:
\beq
\label{eq:4}
\Psi_{\bk}(\br) = \frac{1}{\sqrt{N_{\phi} \nu(\bk)}} \sum_\bm  c_\bm(\br) e^{i \bk \cdot \br_\bm}. 
\eeq
Here $N_{\phi}$ is the number of degenerate states in the LLL, which is equal to the number of magnetic 
flux quanta piercing the sample,  and $\nu(\bk)$ is a normalization factor. 
Assuming Bloch functions are normalized to unity over the sample area, 
the normalization factor is given by:
\beq
\label{eq:5} 
\nu(\bk) = \sum_{\bm} (-1)^{m_1 m_2} e^{-\br_{\bm}^2 / 4 \ell^2} e^{-i \bk \cdot \br_\bm}.
\eeq 
The momentum $\bk$ belongs to the first Brillouin zone (BZ) of the triangular lattice and 
is given by $\bk = k_1 \bb_1 + k_2 \bb_2$, where $\bb_1 = (\hat x - \hat y / \sqrt{3})/a, \,\,
\bb_2 = 2 \hat y/ a \sqrt{3}$ are the basis vectors of the reciprocal lattice. 
Imposing periodic boundary conditions with respect to {\em magnetic translations} 
along the basis directions $\ba_{1,2}$, fixes $k_{1,2}$ to be 
$k_{1,2} = 2 \pi n_{1,2}/ \sqrt{N_{\phi}}$ with integers $n_{1,2}$ satisfying 
$-\sqrt{N_{\phi}}/2 \leq n_{1,2} < \sqrt{N_{\phi}}/2$. 
It follows from Eq.(\ref{eq:3}) that the normalization factor $\nu(\bk)$ vanishes at $\bk = \bk_0$  corresponding to 
$(k_1, k_2) = (\pi, \pi)$. 
As shown in Ref.[\onlinecite{Rashba}], the Bloch function at this momentum is still, however,  well-defined
and can be found by carefully taking the limit  $\bk \rightarrow \bk_0$ in Eq.(\ref{eq:3}). 
Magnetic Wannier functions are then obtained from the Bloch functions by the 
inverse Fourier transform:
\beq
\label{eq:6}
\phi_\bm(\br) = \frac{1}{\sqrt{N_{\phi}}} \sum_{\bk} \Psi_\bk(\br) e^{- i \bk \cdot \br_\bm}. 
\eeq
These functions form a complete orthonormal set of states by construction. 
For further in-depth discussion of the properties of these wavefunctions see Ref.[\onlinecite{Rashba}].

Given the complete orthonormal set of magnetic Wannier functions $\phi_{\bm}(\br)$, we can write down the Hamiltonian 
of interacting bosons, projected to the LLL, using this basis. 
The Hamiltonian has the following general form:
\beq
\label{eq:7}
H = \sum_{\bm_1, \ldots, \bm_4} \langle \bm_1 \bm_2 | V | \bm_3 \bm_4 \rangle b^\dg_{\bm_1} b^\dg_{\bm_2} 
b^\pdg_{\bm_4} b^\pdg_{\bm_3},
\eeq
where $b^\dg_\bm$ creates a boson in a magnetic Wannier state $\phi_\bm(\br)$, and we will assume the 
repulsive interaction between the bosons $V$ to be a contact interaction: $V(\br - \br') = V \delta(\br - \br')$. 
The matrix elements in Eq. (\ref{eq:7}) can be easily evaluated numerically. One finds that all these matrix elements 
are nonzero in the thermodynamic limit and are short-range, in the (imprecise) sense of decreasing in magnitude 
rapidly with the separation between the sites. However, even if one assumes that only the matrix elements between 
nearest-neighbor sites may be retained, one still obtains a very complex Hamiltonian with a lot of distinct 
terms, since the only obvious symmetry Eq.(\ref{eq:7}) possesses is the symmetry of the triangular lattice. 
In its raw form, the magnetic Wannier basis Hamiltonian is then rather useless. 
It turns out, however, that this Hamiltonian does in fact possess a hidden symmetry, which is revealed
in the low-energy long-wavelength limit, in the sense to be defined precisely below. 

\section{Magnetic Wannier basis Hamiltonian in the long-wavelength limit}
\label{sec:3}
To proceed, let us consider our Hamiltonian not in the Wannier but in the  
magnetic Bloch basis, given by Eq.(\ref{eq:4}). 
Explicitly evaluating the matrix element of the contact interaction in the Bloch basis we obtain:
\beq
\label{eq:8}
H = \sum_{\bk, \bq, \bq', \bq''} I(\bk, \bq, \bq', \bq'') \delta_{\bq + \bq', \bq'' + \bG} b^\dg_{\bk + \bq} b^\dg_{\bk + \bq'}
b^\pdg_{\bk + \bq''} b^\pdg_\bk, 
\eeq
where $\bk, \bq, \bq', \bq''$ belong to the first BZ, $\delta_{\bq + \bq', \bq'' + \bG}$ expresses momentum conservation modulo 
a reciprocal lattice vector $\bG$ and the interaction matrix element is given by:
\beqa
\label{eq:9}
&&I(\bk, \bq, \bq', \bq'') \nonumber \\
&=&V \int d \br \Psi^*_{\bk + \bq}(\br) \Psi^*_{\bk + \bq'}(\br) 
\Psi^\pdg_{\bk + \bq''}(\br) \Psi^\pdg_\bk(\br) \nonumber \\
&=&\frac{V / 4 \pi \ell^2}{\sqrt{\nu(\bk + \bq) \nu(\bk + \bq') \nu(\bk + \bq'') 
\nu(\bk)}} \nonumber \\
&\times&\frac{1}{N_{\phi}} \sum_{\bm_1,\bm_2,\bm_3} (-1)^{m_{11}m_{12} + m_{21}m_{22} +
m_{31}m_{32}} \nonumber \\
&\times&e^{-(1/8 \ell^2)[\br_{\bm_1}^2 + \br_{\bm_2}^2 + (\br_{\bm_1} - \br_{\bm_3})^2 
+ (\br_{\bm_2} - \br_{\bm_3})^2]} \nonumber \\
&\times& e^{(i/4 \ell^2) \hat z \cdot [(\br_{\bm_1} + \br_{\bm_2})
\times \br_{\bm_3}]} \nonumber \\
&\times& e^{-i (\bk + \bq) \cdot \br_{\bm_1}} e^{-i (\bk + \bq') \cdot \br_{\bm_2}} 
e^{i (\bk + \bq'') \cdot \br_{\bm_3}}.
\eeqa

We now note the following property of the magnetic Bloch functions. 
Using Eq.(\ref{eq:3}), we can rewrite Eq.(\ref{eq:4}) for the magnetic Bloch function as:
\beq
\label{eq:10}
\Psi_{\bk}(\br) = \frac{1}{\sqrt{N_{\phi} \nu(\bk)}} \sum_\bm  c^*_\bm(\br) e^{i [\bk + (1/\ell^2) \hat z \times \br ] \cdot \br_\bm}. 
\eeq
Using the complex conjugate of the Perelomov overcompleteness identity Eq.(\ref{eq:3}), we then find that the zeros 
of the Bloch function $\Psi_\bk(\br)$ are located at:
\beq
\label{eq:11} 
\br_{\bm \bk} = \br_\bm + \frac{1}{2}\left(\ba_1 + \ba_2 \right) + \ell^2 \hat z \times \bk. 
\eeq
The zeros of the magnetic Bloch functions thus form a triangular lattice, with one magnetic flux 
quantum per unit cell. Different values of the first BZ momentum $\bk$ label different positions 
of this lattice of zeros relative to the lattice formed by the basis magnetic Wannier states. 
Since wavefunctions in the LLL are fully specified, up to phase factors, by their zeros, it follows that the Bloch functions can 
be identified with the Abrikosov vortex lattice states, which form the set of ground states of Eq.(\ref{eq:8}) at 
large filling factors. From the viewpoint of the Hamiltonian in the magnetic Bloch 
basis, the Abrikosov vortex lattice states correspond to condensation of the bosons in states with a particular momentum $\bk$. ``Condensation" here should be understood in the sense of the Bloch states being the 
solutions of the LLL-projected Gross-Pitaevskii equation, which is satisfied by the boson fields $b_\bk$ (which 
become c-numbers in the limit of large filling factor):
\beq
\label{eq:12}
\frac{\partial H}{\partial b_\bk^*} - \mu b_\bk = 0, 
\eeq
where $\mu$ is the chemical potential. 
The solution of this equation, corresponding to the triangular Abrikosov vortex lattice, is given by:
\beq
\label{eq:13}
\mu = 2 I(\bk,0,0,0) |b_\bk|^2, 
\eeq
which determines the filling factor $\nu = |b_\bk|^2/N_{\phi}$ in terms of the chemical potential. 
It will be demonstrated below that $I(\bk, 0,0,0)$ is independent of $\bk$, so that all such solutions describe degenerate
states at the same filling factor, as they should. 
 
The fact that the magnetic Bloch functions correspond to Abrikosov vortex lattice states now leads us to the 
following observation: the functions $\Psi_{\bk}(\br)$ at different $\bk$ must be related to each other 
by magnetic translations. Indeed, we find the following relation:
\beq
\label{eq:14}
\Psi_{\bk}(\br) = e^{i \gamma_\bk} e^{\frac{i}{2} \bk \cdot \br} \Psi_0(\br - \ell^2 \hat z \times \bk). 
\eeq
Here the factor $e^{\frac{i}{2} \bk \cdot \br}$ is an Aharonov-Bohm phase factor from the magnetic translation operator and 
$e^{i \gamma_\bk}$ is given by:
\beqa
\label{eq:15}
e^{i \gamma_\bk}&=&\frac{\Psi_0^*(-\ell^2 \hat z \times \bk)}{\Psi_\bk(0)} = \frac{1}{\sqrt{\nu(0) \nu(\bk)}}  \sum_\bm (-1)^{m_1 m_2}  \nonumber \\
&\times& e^{-(1/4 \ell^2) (\br _\bm + \ell^2 \hat z \times \bk)^2 - \frac{i}{2} \bk \cdot \br_\bm}.
\eeqa
It is important to note that while $\Psi_{\bk + \bG}(\br) = \Psi_{\bk}(\br)$ as it should, 
$e^{i \gamma_{\bk + \bG}} \neq e^{i \gamma_{\bk}}$. 
From Eq.(\ref{eq:14})  it immediately follows that the interaction matrix element in Eq.(\ref{eq:8}) can be written as:
\beq
\label{eq:16}
I(\bk, \bq, \bq', \bq + \bq') = I(0, \bq, \bq',\bq+\bq') f^*(0, \bq, \bq') f(\bk, \bq, \bq'),
\eeq
where all the $\bk$-dependence is contained in the function 
\beq
\label{eq:17}
f(\bk,\bq,\bq') = e^{-i(\gamma_{\bk + \bq} + \gamma_{\bk + \bq'} - \gamma_{\bk + \bq  + \bq'} - \gamma_{\bk})}.
\eeq 
It is clear from the above expressions that all Abrikosov lattice states, corresponding to condensation of the bosons in states with 
different $\bk$,  are degenerate, as they should be. 

Let us now see what the Abrikosov vortex lattice states correspond to in the Wannier basis. 
Transforming the boson creation operator from the Bloch to the Wannier basis:
\beq
\label{eq:18}
b^\dg_\bm = \frac{1}{\sqrt{N_{\phi}}} \sum_\bk b^\dg_\bk e^{i \bk \cdot \br_\bm}, 
\eeq
one can immediately see that the condensation (in the sense defined above) of the bosons into Bloch states
corresponds to states with uniform phase winding along the basis directions of the triangular lattice in the 
Wannier basis, with the phase gradient given by the momentum~$\bk$. 
This nature of the Abrikosov vortex lattice states has important consequences. 

First consequence, that can be seen immediately, is that the imaginary-time action, corresponding to 
long-wavelength boson field phase fluctuations about a given Abrikosov state, will lack the usual 
$({\boldsymbol \nabla} \theta)^2$ term, characteristic of superfluids, since all states with uniform phase 
gradients have the same energy. 
Instead, the action will have the form (after appropriate rescaling of the time and spatial coordinates):
\beq
\label{eq:19}
S \sim \int d \tau d \br \left[ (\partial_\tau \theta)^2 + ({\boldsymbol \nabla}^2 \theta)^2 \right]. 
\eeq
It then follows that the dispersion of small fluctuations around an Abrikosov 
lattice state is quadratic instead of linear (this holds provided the LLL approximation is valid) $\omega \sim \bq^2$. 
This result is well-known and has been obtained before by a number of authors.\cite{Maki,Tesanovic, Moore,Sinova02} 
The above derivation of this result, using magnetic Wannier functions, is probably the simplest and the most physically transparent.
The fact that the excitation spectrum is quadratic, instead of linear, immediately leads one to the conclusion \cite{Sinova02}
that Bose condensation or true off-diagonal long-range order is absent in this system. 
This does not necessarily mean, however, that the system is not superfluid: as was shown in Ref.[\onlinecite{Sinova02}], 
the vortices are still localized at large filling factors and thus the superfluid stiffness is finite.
As the filling factor is reduced, however, one expects a transition from the Abrikosov vortex lattice state (a vortex solid)
into vortex liquid states, some of which will be incompressible quantum Hall liquids.~\cite{Cooper} It is these 
states that are of primary interest to us. 

The second, and the most important consequence for our purposes, is that the lack of the $({\boldsymbol \nabla} \theta)^2$ term  in the 
phase action actually follows from an {\em emergent conservation law}: namely the conservation of the center-of-mass 
of the bosons in any collision process, which becomes exact at long wavelengths. 
To see this, we again return to the expression for the interaction matrix element in the Bloch basis, Eqs. (\ref{eq:16}) and 
(\ref{eq:17}). 
It may seem at first sight that all the $\bk$-dependent phase factors, that appear in Eq.(\ref{eq:17}),
could be removed by a gauge transformation of the boson creation-annihilation operators, i.e. $b_\bk e^{i \gamma_\bk} \rightarrow b_\bk$, 
accompanied by the corresponding redefinition of the Bloch functions $\Psi_\bk(\br)e^{-i \gamma_\bk} \rightarrow \Psi_\bk(\br)$. 
This is, however, generally not possible due to the fact that $e^{i \gamma_\bk}$ does not have the same periodicity in the reciprocal space as the Bloch functions (the whole function $f$, of course, does have the same periodicity
as the Bloch functions). 
To proceed, we will make an approximation: we will assume that we can restrict ourselves to 
configurations of the boson fields, corresponding to long-wavelength distortions of the classical Abrikosov 
lattice ground states. This is certainly a harmless approximation 
in the vortex lattice state itself and should remain harmless even for vortex liquids as long as the vortex lattice correlation length $\xi$ is larger than the magnetic length. 
This is somewhat analogous to the semiclassical nonlinear-sigma-model treatment of  low-dimensional 
quantum antiferromagnets, \cite{Auerbach} which can successfully describe quantum-disordered states in these systems.  
This approximation implies smallness of the excitation momenta 
$\bq, \bq'$ compared to the reciprocal lattice momenta. 
We introduce a cutoff scale $\Lambda$ for the momenta $\bq, \bq'$, so that $|\bq|, |\bq'| < \Lambda$, 
and assume that $\Lambda$ satisfies the inequality  $1/\xi \ll \Lambda \ll 1/\ell$. 
To leading order in the small parameter $\Lambda \ell$ we can then set  $f(\bk, \bq, \bq') \approx 1$ and the interaction matrix element in Eq.(\ref{eq:8}) becomes independent of $\bk$. 
Transforming the Hamiltonian to the magnetic Wannier basis, we obtain Eq.(\ref{eq:7}),
with $\bm$ labeling the magnetic Wannier states and the interaction matrix element given by:
\beqa
\label{eq:20}
\la \bm_1 \bm_2 | V | \bm_3 \bm_4 \ra &=&g_{\Lambda}(\bm_1- \bm_4, \bm_2 - \bm_4) \nonumber \\
&\times&\delta_{\bm_1 + \bm_2, \bm_3 + \bm_4},
\eeqa
where the function $g_{\Lambda}$ depends on the (very loosely defined) momentum cutoff $\Lambda$ 
and its explicit form is thus rather meaningless (in addition, in a more careful derivation this function would be modified by integrating out excitations with 
$|\bq|, |\bq'| > \Lambda$).  
The physically meaningful information in Eq.(\ref{eq:20}) is contained in the Kroenecker delta-symbol, 
which expresses the conservation of the center-of-mass position of the boson pairs, mentioned above,  and is a 
consequence of the approximate independence of the interaction matrix element in (\ref{eq:8}) on $\bk$. 
Such a center-of-mass position conservation, but only in one spatial direction, is obvious and exact in the Landau-gauge orbital basis, where it appears as a direct consequence of the
momentum conservation in the transverse direction.~\cite{Lee} In our formulation, first BZ momentum is no longer exactly conserved (it is conserved up to a reciprocal lattice vector), since 
our choice of the single-particle basis explicitly breaks translational symmetry (this is a price we have to pay for using spatially-localized single-particle states). 
The center-of-mass conservation then becomes an emergent conservation law, which becomes exact at long wavelengths.

\section{Short-range ring-exchange model on triangular lattice}
\label{sec:4}
The center-of-mass conservation law, derived above,  drastically reduces the number of terms in the Wannier basis Hamiltonian. 
The final approximation we will make, the justification for which will be provided below, is that we can 
retain only the shortest-range terms in the Wannier Hamiltonian. 
This is a harmless approximation provided the characteristic range of the matrix element Eq.(\ref{eq:20}), 
which is of order $1/\Lambda$, is much smaller than the vortex lattice correlation length $\xi$. This was precisely the assumption we made in the argument leading to Eq.(\ref{eq:20}) and thus Eq.(\ref{eq:20}) and the above approximation 
are consistent with each other.    

We then arrive at the following simple short-range lattice Hamiltonian on the 
triangular lattice, that we conjecture faithfully represents interacting bosons in the LLL:
\beq
\label{eq:21} 
H = - K \sum_P b^\dag _{\bm_1} b^\dag_{\bm_2} b^\pdg_{\bm_4} b^\pdg_{\bm_3} + U \sum_\bm n^2_{\bm} 
+ \sum_{\bm \bm'} V_{\bm \bm'} n_{\bm} n_{\bm'}.
\eeq
\begin{figure}
\includegraphics[width=4cm]{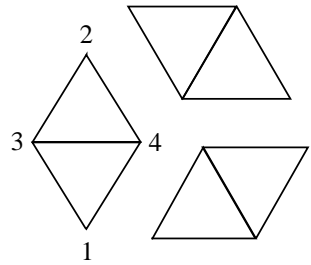}
\caption{Three types of smallest-size 4-site plaquettes $P$ in Eq.(\ref{eq:21}) 
on the triangular lattice. Ring-exchange term hops a pair of bosons on sites $1$ and $2$ to 
sites $3$ and $4$ and back.}
\end{figure}
Eq.({\ref{eq:21}) is the main result of our paper. 
The first term in (\ref{eq:21}) is the shortest-range ring-exhange term on the triangular lattice (there are three distinct kinds of plaquettes $P$, 
as shown in Fig. 1), which is the shortest-range and thus the dominant center-of-mass conserving pair hopping term.  
The second term is on-site repulsion term ($n_{\bm} = b^\dag_{\bm} b^\pdg_{\bm}$). The third term represents 
longer-range repulsion. The relevant range of $V_{\bm \bm'}$ depends on the boson Landau level filling factor $\nu$ 
(i.e. should be at least of the order of the mean interparticle distance for a given filling factor),
and can be restricted to only nearest-neighbor repulsive interactions at $\nu =1/2$.

The sign of $K$ is important and can be fixed by requiring that Eq.(\ref{eq:21}) reproduce 
the correct ground state at large filling factors, i.e. the Abrikosov vortex lattice. 
It is easy to see that at large filling factors the ground state of (\ref{eq:21}) with $K > 0$ 
is in fact the Abrikosov vortex lattice. 
Indeed, in this classical limit we may replace boson operators by c-numbers. The dominant repulsive interaction term in 
Eq.(\ref{eq:21}) is the on-site 
repulsion term. In the classical limit this will favor equal boson density on all the lattice sites. 
The nature of the ground state will then be determined 
by the ring-exchange term, which is the only term in (\ref{eq:21}), that depends on the phases of the bosons and has the form:
\beq
\label{eq:22} 
H = -2 K \sum_P \cos(\theta_{\bm_1} + \theta_{\bm_2} - \theta_{\bm_3} - \theta_{\bm_4}), 
\eeq
where $\theta_\bm$ is the phase of the boson field $b_\bm$.  
It is easy to show \cite{Balents03} that the set of ground states of Eq.(\ref{eq:22}) with $K > 0$ corresponds to all possible uniform phase gradients along the basis directions $\ba_{1,2}$ of the triangular lattice. As already shown in section \ref{sec:3}, this corresponds precisely to Abrikosov vortex lattice 
states. Since all states with uniform phase gradients are degenerate, one also obtains the quadratic dispersion for small phase fluctuations 
around any of the ground states. 
The fact that Eq.(\ref{eq:21}) correctly reproduces both the ground state and the excitation spectrum of the original boson 
Hamiltonian Eq.(\ref{eq:7}) at large filling factors reassures us that it will in fact faithfully represent Eq.(\ref{eq:7}) at all 
filling factors, with properly chosen ring-exchange and interaction parameters. 

In particular, let us now consider the case of the filling factor $\nu=1/2$. For bosons with contact interaction, the exact 
ground state in this case is the $\nu=1/2$ Laughlin liquid:~\cite{Laughlin83}
\beq
\label{eq:23}
\Psi(z_1,\ldots, z_N) = \prod_{i < j} (z_i - z_j)^2 e^{- \sum_i |z_i|^2 / 4 \ell^2}. 
\eeq
By our conjecture that Eq.(\ref{eq:21}) faithfully represents bosons in the LLL, the ground state of this Hamiltonian 
at filling factor $1/2$ is then a featureless Mott insulator with topological order.
We expect this to be true when $K < V$, where $V$ is the strength of the nearest-neighbor repulsive interactions ($U$ is the 
dominant interaction energy scale and can be taken to be large compared to $V$). 
Assuming $U \gg K,V$, the bosons can be taken to be hard-core, i.e. with double occupation of any site of the lattice
prohibited. Using Holstein-Primakoff transformation \cite{Holstein} between hard-core bosons and spins of magnitude $1/2$, we 
can rewrite Eq.(\ref{eq:21}) as the following model of interacting spins-$1/2$ on the triangular lattice:
\beq
\label{eq:24}
H = - K \sum_P S^+_{\bm_1} S^+_{\bm_2} S^-_{\bm_4} S^-_{\bm_3} + V \sum_{\la \bm \bm' \ra} S^z_{\bm} S^z_{\bm'}. 
\eeq
The ground state of Eq.(\ref{eq:24}) at $K=0$ has extensive degeneracy, corresponding to all possible 
configurations of $S^z_i$ with at most one unsatisfied bond per every triangular plaquette of the lattice. ~\cite{Wannier,Houtappel} 
When this degeneracy is lifted by a small two-spin interaction term of the form $-J (S^+_i S^-_j + h.c.)$, 
the ground state is known to be a {\em supersolid}, \cite{supersolid} i.e. a state which has both long-range order 
in the $x,y$-components of the spin and a finite-wavevector ordering of the $z$-components. 
Our mapping between Eq.(\ref{eq:24}) and the $\nu=1/2$ fractional quantum Hall liquid means that the ground 
state of Eq.(\ref{eq:24}), in contrast, is a {\em spin liquid}, i.e. a state with a gapped excitation spectrum and 
topology-dependent ground state degeneracy, which is the same as a featureless Mott insulator in the bosonic language. 
When $K \gg V$, we expect the ground state to be a compressible liquid with a quadratic excitation spectrum,~\cite{Balents03}
most likely a superfluid.  

A somewhat subtle issue, that requires special consideration, is the issue of the ground state degeneracy of the featureless Mott insulator ground 
state of Eq.(\ref{eq:24}) and the nature of its quasiparticle excitations.  The ground state degeneracy 
of the $\nu=1/2$ Laughlin liquid on a torus is 2-fold and the quasiparticles are anyons of charge $\pm e/2$.~\cite{Wen90} 
Both the 2-fold degeneracy and the anyonic nature of the quasiparticles depend crucially on the fact that 
the time-reversal symmetry is broken by the perpendicular magnetic field.\cite{Wen90, Oshikawa} 
However, the Hamiltonians (\ref{eq:21}) and (\ref{eq:24}) are manifestly time-reversal invariant.
The information about the time-reversal symmetry breaking is contained in the Wannier functions $\phi_\bm(\br)$, but not in the center-of-mass 
conserving interaction matrix elements $\la \bm_1 \bm_2 |V| \bm_3 \bm_4 \ra$, which are all real, as can be seen by 
inspection of Eq. (\ref{eq:9}). 
The matrix elements, which do carry the information about the time-reversal breaking, are the ones that do not conserve the 
center-of-mass, as these matrix elements are in general complex. 
The simplest kind of such a matrix element, and also the one that has the largest magnitude at short distances, 
is the ``correlated hopping"-type matrix element with, for example, $\bm_1 = \bm_3,\,\, \bm_2 \neq \bm_4$. 
It is clear that such a matrix element is, in general, complex. It is also very easy to see why such  matrix elements 
are irrelevant at long distances (but see below): one simply needs to notice that
\beq
\label{eq:25}
\sum_{\bm} \langle \bm  \bm_1 | V | \bm  \bm_2 \rangle \sim \delta_{\bm_1, \bm_2}. 
\eeq 
Our main assumption is that this irrelevance continues to hold even at low filling factors, such as $\nu=1/2$. 
This should be true as long as the correlation length in a given state is significantly larger than the magnetic length. 
The only problem with this is that the topological degeneracy on a torus of the incompressible liquid ground state of Eq.(\ref{eq:24}), which is time-reversal invariant,  has to be equal 
to 4 (assuming the quasiparticle charge is $\pm e/2$, as in the Laughlin state), \cite{Oshikawa} i.e. double the degeneracy of the Laughlin liquid. 

The most natural resolution of this apparent paradox seems to be as follows. 
The set of 4 degenerate ground states of Eq.(\ref{eq:24}) must consist of 2 pairs of degenerate 
states, each pair corresponding to the Laughlin liquid with the magnetic field directed along $\hat z$ or 
$- \hat z$, as Eq.(\ref{eq:24}) is invariant under time-reversal. 
It then follows that each such pair of states breaks time-reversal symmetry spontaneously. 
The spin liquid ground state of (\ref{eq:24}) is then a Kalmeyer-Laughlin-type chiral liquid, \cite{Laughlin87}
which spontaneously breaks parity and time-reversal symmetry.~\cite{footnote}
The quasiparticle excitations above such a state are charge $\pm e/2$ anyons, as in the Laughlin liquid. \cite{Kivelson88}
The role of the complex center-of-mass nonconserving matrix elements $\la \bm_1 \bm_2 |V| \bm_3 \bm_4 \ra$, which explicitly break 
time-reversal symmetry, such as the correlated hopping matrix elements mentioned above, is to act as a small
``symmetry-breaking field", that lifts the degeneracy between the 2 pairs of states, but is otherwise unimportant. 
One then obtains a 2-fold degenerate ground state on a torus with anyonic quasiparticle excitations, exactly as in the $\nu=1/2$ Laughlin 
liquid. This scenario is very appealing, especially in light of the fact that there are so far only two examples of microscopic models in 
the literature, which have been shown to have a chiral spin liquid ground state. \cite{Greiter07, Kivelson07}
Both these models, however, are significantly more complicated than Eq.(\ref{eq:24}). 
Our result is of course only a conjecture at this point and needs to be verified by an explicit numerical simulation. 
\section{Conclusions}
\label{sec:5}
In conclusion, we have derived an explicit mapping between the Hamiltonian of interacting bosons 
in the LLL and a time-reversal invariant Hamiltonian of interacting bosons on the 
triangular lattice with one flux quantum per unit cell, Eq.(\ref{eq:21}). 
At the filling factors, at which the bosons in the LLL condense into incompressible quantum Hall liquid states (such as $\nu=1/2$), the ground 
state of this lattice Hamiltonian is a featureless Mott insulator with topological order and spontaneously broken 
time-reversal symmetry. 
The ground state degeneracy of the featureless Mott insulator state on a torus is thus predicted to be equal to twice the 
ground state degeneracy of the corresponding Laughlin state, i.e. 4 in the case of filling factor $1/2$. 
By the same logic, at odd-denominator filling factors, such as $\nu=1/3$, the ground states of (\ref{eq:21}) should be compressible 
but non-superfluid liquids (``Bose metals"),~\cite{MPAF} corresponding to composite fermion Fermi liquid ground states of 
2D bosons in magnetic field.~\cite{Halperin93}
All these predictions are testable by either quantum Monte-Carlo simulations, since Eq.(\ref{eq:21}) does not have a sign problem, or by exact diagonalization of~(\ref{eq:21}). 
While we have demonstrated the FQHL to featureless Mott insulator connection for the case of interacting bosons, 
we believe that our conclusions also hold, with possible minor modifications, in the case of interacting fermions as well, 
since the physics of the FQHE and of Mott insulators does not depend significantly on the statistics of the~particles.

\begin{acknowledgments}
I thank A.H. MacDonald, R.G. Melko, and especially 
A. Paramekanti for useful discussions. Financial support was provided by the NSERC of Canada and a University of Waterloo start-up grant.
\end{acknowledgments}


\begin{thebibliography}{99}
\bibitem{footnote} The claim of broken time-reversal symmetry will perhaps appear less ``exotic" in this context 
if one notices that the Abrikosov vortex lattice states in the magnetic Wannier description also in general
spontaneously break time-reversal, as they correspond to condensation into states with a definite Bloch momentum.
\end{thebibliography}
\end{document}